Technical Note

# On stepwise advancement of fractures and pressure oscillations in saturated porous media

C. Peruzzo[a], L. Simoni[b], B.A. Schrefler[b,c]

[a] EPFL (Ecole Polytechnique Federale de Lausanne), Geo Energy Laboratory, EPFL-ENAC-IIC-GEL 1015 Lausanne, Switzerland
[b] Department of Civil, Environmental and Architectural Engineering, University of Padua, via Marzolo, 9, Padova, Italy
[c] Institute for Advanced Study, Technische Universität München, Lichtenbergstrasse 2a, D-85748 Garching b. München, Germany

We refer to Fig. 23 and the related comment of the paper K.M. Pervaiz Fathima, René de Borst, Implications of single or multiple pressure degrees of freedom at fracture in fluid saturated porous media, Engineering Fracture Mechanics, 213 (2019), 1–20 (see Fig. 1).

The comment to Figure 23 is "Fig. 22a also shows an increase in crack length which seems to be stepwise [31]. While it is physically quite well possible that such a phenomenon exists, and there is evidence from simulations using discrete models [32], it may be less likely that a continuum model can predict such a discrete phenomenon. Indeed, Fig. 23 suggests that the stepwise advancement disappears upon refinement of the discretization and the time step." (References and Figure numbers of Fathima and De Borst, 2019). This statement is misleading.

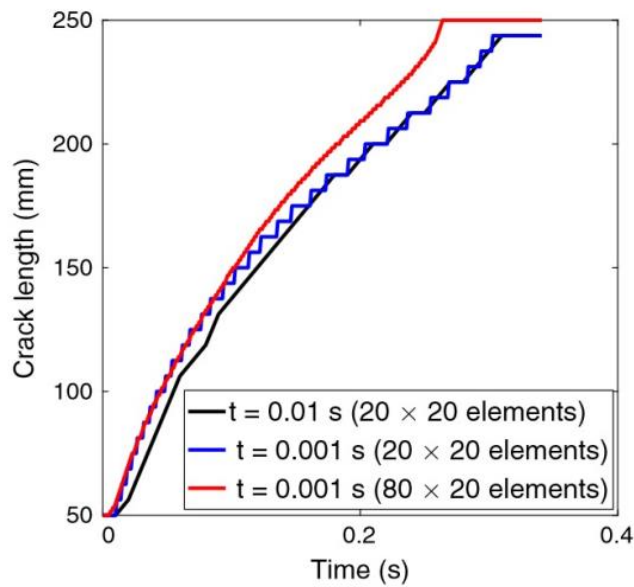

**Figure 1**, taken from Fathima and De Borst, 2019

First of all we want to stress that stepwise fracture tip advancement and related pressure oscillation in saturated porous media are not only "quite well possible" but do exist and have duly been documented with experiments in [1–6] and with field observations in [7–12]. Intermittent fracture advancement in saturated formations is also known from geophysical observations [13–23]. Without such a behaviour, the non-volcanic (subduction) tremor and volcanic tremor are difficult to explain.

The phenomenon has important economic relevance in fracking operations [11] and it would be a pity if computational continuum models would not be able to simulate it properly. But this is not the case as shown in [9,24–34]. These authors used either standard Finite Elements, XFEM, Finite Volume, Finite Differences on originally continuum models. All solutions feature irregular advancement steps which point to a physical origin. Regular steps on the contrary are mainly of numerical origin. In fact Figure 23 reproduces exactly what was evidenced in [35]: in a fracturing problem solved with XFEM in [36] advancement with regular steps can clearly be observed in the crack length histories in Figure four of that paper. Upon use of finer discretization both in space and time these steps disappeared [37]. In fact on two hydro-fracturing problems solved with a finer mesh in [38] smooth solutions were obtained as in Figure 23. This just means that the time step/fracture advancement algorithm employed is not able to catch the phenomenon because it interferes with the three velocities which matter: crack tip advancement velocity and that of the fluid in the fracture and in the surrounding region of the crack tip. Pressure oscillations have then be obtained in [39] by the senior author of the above two papers [36,38] together with co-authors where it is stated that "…fluctuations occur as a result of hydro-fracture evolution" and "Notably the current framework is capable of capturing the addressed fluctuations as it is apparent from the numerical simulation results". All just depends on the time step/fracture advancement algorithm employed: an improper algorithm yields regular steps. The problem referred to in [35,36] is that of the rupture of a saturated square plate (0.25 · 0.25 m) in plane strain conditions under a prescribed fixed vertical velocity $v = 2.35 \cdot 10^{-2}$ μm/s in the opposite direction at the top and bottom of the plate (tensile loading), originally proposed in [40] where no advancement was shown. It has been solved again in [41] both with standard Galerkin Finite Elements and a central force lattice model and the staccato tip advancement with irregular steps was confirmed. The problem evidenced in Fig. 1 and discussed above points to a consistency issue where the influence of the time step/fracture advancement algorithm has been ignored. This is now under further investigation.

A further proof of stepwise advancement is given by forerunning of cracks. Forerunning exists in rocks [42] and can be obtained with the double cantilever beam of [43] as shown on a longer sample than in [43] (forerunning not shown there) in Figs. 2 and 3 where a single forerunning event can be clearly recognized. The external load $q(x,t)$ is monotonically increasing in time as expressed by the following:

$$q(x,t) = \frac{1}{2\pi(x+0.9)} \arctan\left(\frac{t^4}{2}\right)$$

The constitutive law for the cohesive zone is expressed as follows:

$$f(v) = \Gamma(-L_0 + x) \cdot [1 - \Gamma(-v_{max} + v)] \cdot [k\Gamma(v) + \beta\Gamma(-v)] \cdot v$$

where Γ(ξ)=0 if ξ<0 and Γ(ξ)=1 otherwise, (see [43]). β is a large parameter used within a penalty method when the fracture closes.

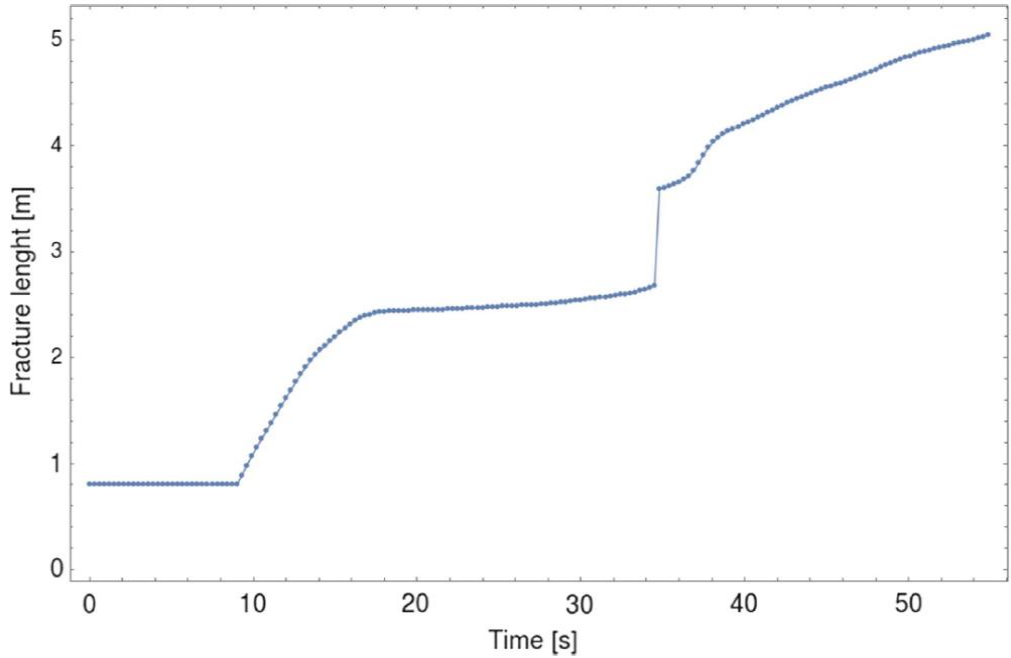

**Figure 2.** Fracture length vs time. Adopted parameters: $L_0 = 0.8$, $L = 10$, $\rho A = 500$, $EJ = 0.15$, $k = 18$ as reported in [43].

The initial notch $L_o = 0.8$ m is present since the beginning of the simulation. In snapshot one the real fracture apex and the cohesive zone are evidenced. At $t \sim 10$ s, Fig. 2, the fracture starts growing and then decelerating, ($t \sim 10$ s to $t \sim 35$ s), until a very small value of the velocity is reached. At $t \sim 35$ s a sudden fracture increment takes place (snapshots 3 and 4 in Fig. 3).

The behaviour is similar to what shown for a single cluster in [44] where for an incident sinusoidal wave forerunning events repeat themselves periodically. Each time the primary crack catches up with the forerunning crack there appears a big jump in the crack advancement. If on top one thinks of a fluid in the crack, clearly pressure fluctuations are expected because of the large volume changes in the fracture. There is no reason whatsoever that this cannot be obtained by Finite Elements. In fact some sort of forerunning has been obtained in [45] with a FE model on a plane grid for a single phase sample. In this paper is stated: "Sometimes a secondary crack develops ahead of the main crack and subsequently links up with it. This leads to a situation where the crack appears to be stationary for a relatively long time and then grows rather abruptly".

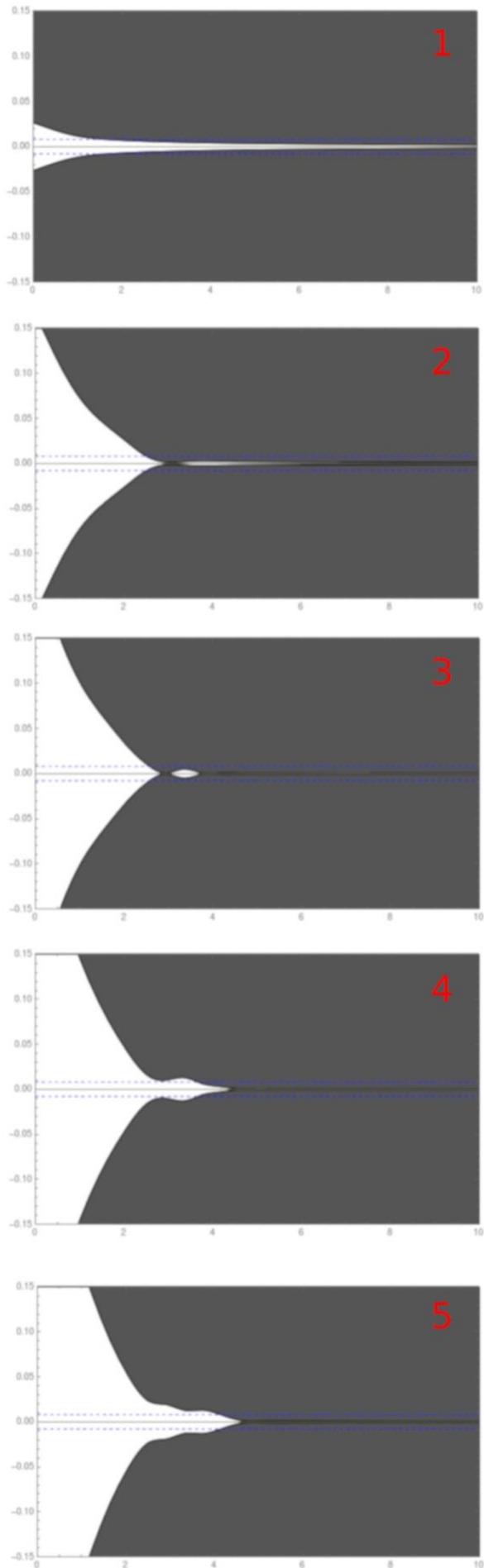

**Figure 3.** From 1 to 5, snapshots sequence of the system evolution. The cohesive zone is active where the fracture opening (2v) is less than the distance between the dashed lines ($2v_{max}$). The sudden length increase happens between fig. 3 and 4.

time

In conclusion, there is no doubt that stepwise fracture advancement exists and that it can be obtained with continuum models provided that a proper time step/fracture advancement algorithm is employed which interferes as little as possible with the velocities at play.

## Acknowledgements

The authors were supported by the project 734370-BESTOFRAC ("Environmentally best practices and optimization in hydraulic fracturing for shale gas/oil development")-H2020-MSCA-RISE-2016.